\documentclass[12pt]{article}
\usepackage{graphicx}
\begin{document}
\begin{center}   {Instantaneous Power Radiated from Magnetic Dipole Moments}
\end{center}
\vspace{.2in}
\begin{center}
P.D. Morley\footnote{{\it E-mail address:} peter3@uchicago.edu} \\  
System of Systems Analytics, Inc. \\ 11250 Waples Mill Road \\ Suite 300 \\Fairfax, VA 22030-7400 \\
and\\
D.J. Buettner\footnote{{\it E-mail address:} Douglas.J.Buettner@aero.org} \\
Aerospace Corporation\\P. O. Box 92957\\Los Angeles, CA 90009-2957
\end{center}

\begin{abstract}
We compute the power radiated per unit solid angle of a moving magnetic dipole moment, and its instantaneous radiated power, both non-relativistically and relativistically. This is then applied to various interesting situations: solar neutrons, electron synchrotrons and cosmological Dirac neutrinos. Concerning the latter, we show that hypothesized early-universe Big Bang conditions allow for neutrino radiation cooling and provide an energy loss-mechanism for subsequent neutrino condensation.
\end{abstract}
  
keywords: neutrinos, dipole radiation, condensation

\section{Introduction}
\hspace{0.25in}The total instantaneous  radiated power\cite{Jac} of a particle carrying charge Q is the Larmor formula for non-relativistic motion
\begin{equation}
P_{QNR} = \frac{2}{3}\frac{Q^{2}}{c^{3}}|\dot{\vec{v}}|^2
\end{equation}
and the Li\'{e}nard formula for relativistic motion
\begin{equation}
P_{QR} =\frac{2}{3}\frac{Q^{2}}{c}\gamma^{6}[(\dot{\vec{\beta}})^{2} -(\vec{\beta} \times \dot{\vec{\beta}})^{2}]
\end{equation}

In this paper, we use Gaussian units with $c$ the speed of light, $\gamma$ the usual Lorentz factor $1/ \sqrt{1 - (\vec{\beta})^{2}}$, dots $\dot{}$ signifying time-derivatives and $\vec{\beta} \equiv \vec{v}/c$, where $\vec{v}$ is the particle's velocity. Also, the neutrinos considered here are Dirac neutrinos with magnetic dipole moments $\vec{\mu} \neq 0$. In this paper, $\vec{\mu}$ will always denote rest frame magnetic moments. Using the same convention as \cite{Jac}, the electromagnetic field notation will be: unprimed quantities such as $\vec{B}$ denotes laboratory quantities (here a laboratory magnetic field) and the $\prime$, such as $\vec{B^{\prime}}$, denotes rest frame quantities (here a rest frame magnetic field). We will find that the corresponding formulae for radiation power losses of magnetic dipoles are much more complicated than Eq(1,2) of charged particles because 

\begin{equation}
\frac{dQ}{dt} = 0 \; {\rm (charge \; conservation)}
\end{equation}
\begin{equation}
\dot{\vec{\mu}} \equiv \frac{ d \vec{\mu}}{dt} \neq 0 \; {\rm (spin \; precession)}
\end{equation}

In the early universe, once cosmological neutrinos decouple from matter, conventional wisdom posits that they would then lose energy adiabatically from the cosmic expansion \cite{Kolb}. However, neutrinos will still interact with early-universe magnetic fields $\vec{B}$. Such primordial fields are thought to be non-uniform (fluctuating) and large $|\vec{B}| > 4.4 \times 10^{13}$ G \cite{Gra, Wag}. We will show that these early-universe magnetic field conditions can allow efficient neutrino cooling after they decouple from matter.

\section{Non-relativistic Instantaneous Radiated Power}
\hspace{0.25in}Reference \cite{Jac} gives the energy radiated per unit solid angle per unit frequency interval for a moving magnetic dipole moment ($\vec{n}$ is the look unit vector): 
\begin{equation}
\frac{d^{2}I}{d \omega d \Omega} = \frac{\omega^{4}}{4 \pi^{2} c^{3}} | \int dt \; \vec{n} \times [\vec{\mu} + \vec{n} \times
(\vec{\beta} \times \vec{\mu})]e^{i\omega (t - \vec{n} \cdot \vec{r}(t)/c)}|^{2}
\end{equation}

Integration by parts yields the time-dependent vector potential $\vec{A}(t)$
\begin{equation}
\vec{A}(t) = \frac{1}{\sqrt{4 \pi c^{3}}} \frac{d^{2}}{dt^{2}} \{ \hat{n} \times \vec{\mu} + \hat{n} \times [ \hat{n} \times (\vec{\beta} \times \vec{\mu})] \}
\end{equation}
Doing the differentiation, we get our first result, which is the non-relativistic angular distribution
\begin{equation}
\frac{dP_{MNR}}{d \Omega} = \frac{1}{4 \pi c^{3}} | \hat{n} \times \ddot{\vec{\mu}} + \hat{n} \times [ \hat{n} \times [ \ddot{\vec{\beta}} \times \vec{\mu}  + 2\dot{\vec{\beta}} \times \dot{\vec{\mu}} +
 \vec{\beta} \times \ddot{\vec{\mu}}]]|^{2}
\end{equation}

Interestingly, the `jerk' $\ddot{\vec{\beta}}$ contributes, which is not the case for an accelerated charge. Altogether, there are 10 separate terms (compare to Eq(1) above) for the instantaneous radiation of a non-relativistically moving magnetic dipole, obtained by integrating Eq(7)

\begin{equation}
P_{MNR} = \frac{2}{3 c^{3}} [\ddot{\vec{\mu}}^{2} + (\ddot{\vec{\beta}} \times 
\vec{\mu} + 2 \dot{\vec{\beta}} \times \dot{\vec{\mu}} + \vec{\beta} \times \ddot{\vec{\mu}})^{2}]
\end{equation}

Eq(8) will be applied to solar neutrons to see if magnetic dipole moment radiation is a factor in their detection in the inner parts of the solar system.

\section{Non-relativistic Spin and Force Equations}
\hspace{0.25in}We are interested in solar neutrons associated with large flares. A typical energy is 8-10 MeV \cite{Share}. The neutrons can be modeled as originating near the solar limb and going through a high magnetic field $\sim$ 10$^{4}$ G (i.\ e. about 1 Tesla) where such strength extends to about one solar diameter away. The question is whether or not these neutrons radiate energy away before they leave the sun's environment. Because of the tenuous low density plasma that they transit and because of their low interaction nuclear cross section, solar neutrons can be considered to be a collisionless particle. However, the neutron will experience forces due to the solar magnetic field.

The force equation is 
\begin{equation}
\vec{F} = \vec{\nabla}(\vec{\mu} \cdot \vec{B^{\prime}} )
\end{equation}
where, as already noted, $\vec{\mu}$ and $\vec{B^{\prime}}$ are rest frame fields. 

Working this out gives

\begin{equation}
m\ddot{\vec{x}} = (\vec{\mu} \cdot \vec{\nabla})\vec{B^{\prime}} + (\vec{B^{\prime}} \cdot \vec{\nabla})\vec{\mu} + \vec{\mu} \times (\vec{\nabla} \times \vec{B^{\prime}}) + \vec{B^{\prime}} \times (\vec{\nabla} \times \vec{\mu})
\end{equation}

We need the jerk, so we have to take the time derivative of Eq(10) to obtain:

\begin{eqnarray}
m \stackrel{...}{\vec{x}}   =    (\dot{\vec{\mu}} \cdot \vec{\nabla})\vec{B^{\prime}} + (\vec{\mu} \cdot \vec{\nabla})\dot{\vec{B^{\prime}}}  +  (\dot{\vec{B^{\prime}}}  \cdot \vec{\nabla})\vec{\mu} +  (\vec{B^{\prime}} \cdot \vec{\nabla})\dot{\vec{\mu}} \nonumber \\
 + \dot{ \vec{\mu}} \times (\vec{\nabla} \times \vec{B^{\prime}}) +  \vec{\mu} \times (\vec{\nabla} \times \dot{ \vec{B^{\prime}}}) + \dot{\vec{B^{\prime}}} \times (\vec{\nabla} \times \vec{\mu}) + \vec{B^{\prime}} \times (\vec{\nabla} \times \dot{\vec{\mu}})
\end{eqnarray}

The next vector equation is the torque equation (no collisional relaxation term present and since we take $\vec{v} \sim 0$)
\begin{equation}
\vec{\mu} \times \vec{B^{\prime}} = \frac{ d \vec{s}}{d t}
\end{equation}
where $\vec{s}$ is the spin of the neutron. The connection between the spin $\vec{s}$ and the magnetic moment $\vec{\mu}$ is $ \vec{\mu} = u \vec{s}$ where $u$ is a constant. So the third vector equation is
\begin{equation}
\dot{\vec{\mu}} = u \vec{\mu} \times \vec{B^{\prime}}
\end{equation}

Finally, we need $\ddot{\vec{\mu}}$ given by taking the time derivative of Eq(13):
\begin{equation}
\ddot{\vec{\mu}} = u^{2}\{ (\vec{B^{\prime}} \cdot \vec{\mu}) \vec{B^{\prime}} - \vec{B^{\prime}}^{2} \vec{\mu} \} + u \vec{\mu} \times \dot{\vec{B^{\prime}}}
\end{equation}

\section{Solar Neutron Energy Loss}
\hspace{0.25in}The neutron magnetic moment \cite{PDG} is $\mu_{Neu} = -1.9130427 \mu_{N}$ where $\mu_{N} = \frac{e \hbar}{2 m_{P}c}$ is the nuclear magneton. If we assume that the solar magnetic field has small variations and fluctuations, then only the first term in Eq(8) contributes, with $\ddot{\vec{\mu}}$ given by Eq(14), which has the maximum value
\begin{equation}
|\ddot{\vec{\mu}}|^{2} \leq (u^{2} \vec{B^{\prime}}^{2} \mu_{Neu})^{2}
\end{equation}
The space (`labatory') frame of one Tesla for the sun's environment translates to a higher strength value in the neutron's rest frame by a $\gamma$. Since the neutrons are non-relativistic ($\gamma \sim 1$), we neglect any Lorentz transformation involved. Working this out for a nominal one Tesla $\vec{B^{\prime}}$, we find
\begin{equation}
P^{neutron}_{NR} \sim 1.625 \times 10^{-33} \; {\rm eV/sec}
\end{equation}
Even if we include a non-stationary solar magnetic field component (therefore other terms in Eq(8) come into play), neutron radiation losses are completely negligible for this problem.

\section{Relativistic Instantaneous Radiated Power}
\hspace{0.25in}From the last section, it is obvious that non-negligible magnetic moment radiation requires great magnetic field strengths and ultra-relativistic particle motion. $P_{MNR}$, Eq(8), in section 2 and the torque equation, Eq(12), and force equation, Eq(9), both in section 3, all have to be generalized. In this section, we present the generalization of the instantaneous power radiated.

The instantaneous power is a Lorentz scalar, so time is replaced with proper time, a scalar. Lorentz tensors come into play, but they have to be contracted together to get a scalar.

The formula clearly has two separate quantities reflecting the two separate terms in the non-relativistic form. The first term is the double time derivative of the magnetic moment. Magnetic moments are entries of the anti-symmetric dipole tensor $D_{\mu \nu}$ \cite{PEL}
\begin{equation}
D_{\mu\nu} = \left( \begin{array}{cccc} 0 & d_{1} &  d_{2} & d_{3} \\ - d_{1} &  0 & \mu_{3} & -\mu_{2} \\ -d_{2} & -\mu_{3} &  0 &  \mu_{1} \\ -d_{3} & \mu_{2} & -\mu_{1}  & 0 \end{array} \right)
\end{equation}
where $\vec{\mu}$ is the magnetic moment vector and $\vec{d}$ is the electric dipole vector, both in the rest frame of the particle. Experimentally, no elementary particle has been found (Particle Data Group \cite{PDG}) that carries an intrinsic (rest frame) non-zero $\vec{d}$ so we set this term to zero. 

The first term thus becomes
\begin{equation}
\frac{2}{3 c^{3}} [\ddot{\vec{\mu}}^{2}] \rightarrow \frac{2}{3 c^{3}} [\frac{1}{2} \frac{d^{2} D^{\mu\nu}}{d \tau^{2}} \frac{d^{2} D_{\mu\nu}}{d \tau^{2}}]
\end{equation}

Using
\begin{equation}
\frac{d}{d \tau} = \gamma \frac{d}{d t}
\end{equation}
and
\begin{equation}
\frac{d \gamma}{d t} = \gamma^{3} \vec{\beta} \cdot \dot{\vec{\beta}}
\end{equation}
we get
\begin{equation}
\frac{2}{3 c^{3}} [\ddot{\vec{\mu}}^{2}]   \rightarrow   \frac{2}{3 c^{3}} [\gamma^{8} (\vec{\beta}\cdot \dot{\vec{\beta}})^{2} \dot{\vec{\mu}}^{2}  
 + 2 \gamma^{6}(\vec{\beta}\cdot \dot{\vec{\beta}})( \dot{\vec{\mu}}\cdot \ddot{\vec{\mu}}) + \gamma^{4}\ddot{\vec{\mu}}^{2}]
\end{equation}

The remaining is the piece
\begin{equation}
\frac{d^{2}(\vec{\beta} \times \vec{\mu})}{d t^{2}} = \ddot{\vec{\beta}} \times \vec{\mu} + 2 \dot{\vec{\beta}} \times \dot{\vec{\mu}} + \vec{\mu} \times \ddot{\vec{\mu}}
\end{equation}
so the second and final term in the instantaneous power emitted is 
\begin{equation}
\frac{2}{3 c^{3}} [ (\ddot{\vec{\beta}} \times \vec{\mu} + 2 \dot{\vec{\beta}} \times \dot{\vec{\mu}} + \vec{\beta} \times \ddot{\vec{\mu}})^{2}] \rightarrow \frac{2}{3 c^{3}} [\frac{1}{m^{2}c^{2}}
\frac{d^{2}(D^{\nu\mu}P_{\mu})}{d \tau^{2}} \frac{d^{2}(D_{\nu\lambda}P^{\lambda})}{d \tau^{2}}]
\end{equation}
Hence, the relativistic instantaneous power $P_{MR}$ radiated from a magnetic dipole moment is
\begin{equation}
P_{MR} = \frac{2}{3 c^{3}} [\frac{1}{2} \frac{d^{2} D^{\mu\nu}}{d \tau^{2}} \frac{d^{2} D_{\mu\nu}}{d \tau^{2}} +   \frac{1}{m^{2}c^{2}}
\frac{d^{2}(D^{\nu\mu}P_{\mu})}{d \tau^{2}} \frac{d^{2}(D_{\nu\lambda}P^{\lambda})}{d \tau^{2}}]
\end{equation}
Working out the algebra gives the final answer
\begin{eqnarray}
P_{MR} &  = &   \frac{2}{3 c^{3}} [\gamma^{8} (\vec{\beta}\cdot \dot{\vec{\beta}})^{2} \dot{\vec{\mu}}^{2}  
 + 2 \gamma^{6}(\vec{\beta}\cdot \dot{\vec{\beta}})( \dot{\vec{\mu}}\cdot \ddot{\vec{\mu}}) + \gamma^{4}\ddot{\vec{\mu}}^{2}  \nonumber \\
&  + &  \gamma^{8}(\vec{\beta}\cdot \dot{\vec{\beta}})^{2}(\dot{\vec{\beta}} \times \vec{\mu} + \vec{\beta} \times \dot{\vec{\mu}})^{2} + 2 \gamma^{6}(\vec{\beta} \cdot \dot{\vec{\beta}})(\dot{\vec{\beta}} \times \vec{\mu}+ \vec{\beta} \times \dot{\vec{\mu}}) \cdot \nonumber \\
& &  (\ddot{\vec{\beta}} \times \vec{\mu} + 2 \dot{\vec{\beta}} \times \dot{\vec{\mu}} + \vec{\beta} \times \ddot{\vec{\mu}}) + \gamma^{4}(\ddot{\vec{\beta}} \times \vec{\mu} + 2 \dot{\vec{\beta}} \times \dot{\vec{\mu}} + \vec{\beta} \times \ddot{\vec{\mu}})^{2} ]
\end{eqnarray}

Eq(25) is the relativistic magnetic dipole moment radiation loss equivalent to the  Li\'{e}nard formula Eq(2) for charged particles. 

\section{Electron Synchrotrons}
\hspace{0.25in}Eq(25) can be applied immediately to electron synchrotrons, using an example, the Cornell electron storage ring \cite{COR}.
Electrons will radiate magnetic dipole radiation in conjunction with their usual charge radiation. Since the machine designers have not accounted for the former, it must be truely tiny, even for large accelerators. 

For circular machines (call the loss $P_{C}$), the energy change per revolution is small, so $\vec{\beta} \cdot \dot{\vec{\beta}} \rightarrow 0$, and 
since $\dot{\vec{\mu}}$,  $\ddot{\vec{\mu}} \simeq 0$, the only term in Eq(25) that is present is
\begin{equation}
P_{C} = \frac{2}{3 c^{3}} \gamma^{4} [\ddot{\vec{\beta}} \times \vec{\mu}]^{2}
\end{equation}
Now
\begin{equation}
|\ddot{\vec{\beta}}| = \omega |\dot{\vec{\beta}} |
\end{equation}
and assuming perpendicularity between $\ddot{\vec{\beta}}$ and $\vec{\mu}$
\begin{equation}
P_{C} = \frac{2}{3 c^{3}} \gamma^{4} \vec{\mu}^{2} \omega^{2} |\dot{\vec{\beta}} |^{2}
\end{equation}
The acceleration $|\vec{a}| = \frac{v^{2}}{r} \simeq \frac{c^{2}}{r}$ where $r$ is the radius of the synchrotron. Thus 
$|\dot{\vec{\beta}} | \cong \frac{c}{r}$. With $\omega = \frac{c}{r}$, the radiative energy loss per revolution $\delta E$ is 
\begin{equation}
\delta E = \frac{2 \pi r}{c} P_{C} = \frac{4 \pi}{3} \gamma^{4} \vec{\mu}^{2}/r^{3}
\end{equation}
Expressing the energy loss per revolution in $MeV$, $\delta E(MeV)$, we find
\begin{equation}
\delta E(MeV) = 1.328 \times 10^{-32} \frac{E[Gev]^{4}}{r[meters]^{3}}
\end{equation}
where the beam energy $E[GeV]$ is expressed in $GeV$ and the accelerator radius $r[meters]$ is expressed in meters. For the Cornell machine, $E[Gev] \simeq 12$ and $r[meters] \simeq 122$ giving
\begin{equation}
\delta E(MeV) = 1.51 \times 10^{-34}
\end{equation}
which, indeed, is completely ignorable.

\section{The Generalized Neutron/Neutrino Force Equation}
\hspace{0.25in}Though the solar neutron problem and the electron synchrotron problem have vanishingly small radiative dipole power losses, we will see that cosmological neutrinos in the early universe may dramatically cool down from this mechanism. To do the calculation, we must have the generalization of the non-relativistic force equation and the non-relativistic torque equation. Of the two generalizations, the former is trivial and the latter complicated.

Introducing the velocity 4-vector $U^{\mu}$
\begin{equation}
U^{\mu} = (\gamma c, \gamma \vec{v}), \; U_{\mu} = (\gamma c, -\gamma \vec{v})
\end{equation}
and the gradient 4-vector $\partial^{\mu}$
\begin{equation}
\partial^{\mu} = (\frac{\partial}{\partial c t}, -\nabla)
\end{equation}
the generalization of Eq(9) is
\begin{equation}
m\frac{d U^{\mu}}{d \tau} = \partial^{\mu}Q
\end{equation}
where $m$ is the invariant rest mass and
\begin{equation}
Q = \frac{1}{2} F^{\alpha \beta \prime}D_{\alpha \beta} = -\vec{\mu} \cdot \vec{B}^{\prime} = -\vec{\mu}\cdot [\gamma(\vec{B} - \vec{\beta} \times \vec{E}) -\frac{\gamma^{2}}{1+\gamma} \vec{\beta}(\vec{\beta} \cdot \vec{B})]
\end{equation}
We will need the expression $\frac{d \vec{\beta}}{d \tau}$ so reducing Eq(34) gives
\begin{equation}
\frac{d \vec{\beta}}{d \tau} = \frac{-\vec{\beta}}{\gamma m c^{2}} \dot{Q} - \frac{\vec{\nabla} Q}{\gamma m c}
\end{equation}

\section{The Generalized Neutron/Neutrino Moment Equation}
\hspace{0.25in}We have to generalize Eq(12) to make it covariant. The resulting equation for charged particles\cite{Jac} is the Thomas\cite{Th}-Bargmann\cite{Ba}-Michel-Teledi equation. For want of a better name, the neutron/neutrino equation we need will just be called the neutrino moment (NM) equation.  We follow \cite{Jac}.

The relativistic spin in the rest frame will be denoted by $\vec{s}$ and the spin in the laboratory frame denoted by $S^{\alpha}$, where this relativistic spin is an axial 4-vector. Recall that spin is an intrinsic quantum operator where the value of $g$ in $\vec{\mu} = \frac{ge}{2 mc} \vec{s}$ has a quantum origin. In particular, the quantum 3-spin $\vec{\Sigma}$ satisfies $[\Sigma^{k}, \Sigma^{l}] = i \epsilon^{klm}\Sigma^{m}$. Because the quantum origins of $g$ may preclude a direct evaluation from first principles (e.\ g. the anomalous g value of the nucleons), the relativistic formalism finesses the possible unknown value of the g term, by making it a parameter. The relationship between $\vec{s}$ and $\vec{S}$ is \cite{Jac}
\begin{equation}
\vec{s} = \vec{S} - \frac{\gamma}{1+\gamma}(\vec{\beta} \cdot \vec{S})\vec{\beta}
\end{equation}
where $S^{\alpha}$ satisfies
\begin{equation}
U_{\alpha}S^{\alpha} = 0
\end{equation}
As given in \cite{Jac}, the possible covariant terms for the time evolution of $S^{\alpha}$ is
\begin{equation}
\frac{d S^{\alpha}}{d \tau} = A_{1}F^{\alpha \beta}S_{\beta} + \frac{A_{2}}{c^{2}}(F^{\lambda \mu}S_{\lambda}U_{\mu})U^{\alpha} + \frac{A_{3}}{c^{2}}(S_{\beta}\frac{d U^{\beta}}{d \tau})U^{\alpha}
\end{equation}
Multiplying Eq(39) by $U_{\alpha}$ and using Eq(38), we determine that $A_{1} = A_{2}$ and $A_{3} = -1$. Since Eq(39) must reduce to Eq(12) when $\vec{v} = 0$, then $A_{1} = \frac{ge}{2mc}$ and we obtain the covariant equation
\begin{equation}
\frac{d S^{\alpha}}{d \tau} = \frac{ge}{2mc}[F^{\alpha \beta}S_{\beta} + \frac{U^{\alpha}}{c^{2}}(F^{\lambda \mu}S_{\lambda}U_{\mu})] -\frac{U^{\alpha}}{c^{2}}(S_{\lambda}\frac{d U^{\lambda}}{d \tau})
\end{equation} 
We process this equation, eventually obtaining an equation for the $\frac{d\vec{\mu}}{dt}$
\begin{equation}
\frac{d \vec{s}}{dt} = \frac{ge}{2mc} \vec{s} \times [\vec{B} - \frac{\gamma}{1+\gamma}(\vec{\beta}\cdot \vec{B})\vec{\beta} -\vec{\beta}\times\vec{E}] +\frac{\gamma}{1+\gamma}[\vec{s} \times (\vec{\beta} \times \frac{d \vec{\beta}}{d \tau})]
\end{equation}
finally giving the NM equation
\begin{equation}
\frac{d \vec{\mu}}{dt} = u \vec{\mu}  \times [\vec{B} - \frac{\gamma}{1+\gamma}(\vec{\beta}\cdot \vec{B})\vec{\beta} -\vec{\beta}\times\vec{E}] +\frac{\gamma}{1+\gamma}[\vec{\mu} \times (\vec{\beta} \times \frac{d \vec{\beta}}{d \tau})]
\end{equation}
where $\frac{d \vec{\beta}}{d \tau}$ is given by Eq(36) in conjunction with Eq(35). In Eq(42), $u$ is the constant in $\vec{\mu} = u \vec{s}$ so
\begin{equation}
u = \frac{g e}{2 mc}
\end{equation}

\section{Cosmological Neutrino Energy Loss}
\hspace{0.25in}We already know that non-negligible magnetic dipole radiation loss from individual elementary particles requires magnetic field strengths comparable to those
found in neutron stars. Early universe cosmological neutrinos that decouple from matter at temperatures $kT \simeq$ 2.6 MeV \cite{Wag} have energy $\frac{3}{2}kT \simeq$ 3.9 MeV giving a $\gamma \approx 3.9 \times 10^{6}$ (since $m_{\nu} \sim 1$ eV) with fluctuating magnetic field strengths $|\vec{B}| > 4.4 \times 10^{13}$ G \cite{Gra, Wag}. In this section, we estimate their power loss after they decouple from matter.

The key is the existence of the $\gamma^{8}$ terms of Eq(25). For decoupling neutrinos, with $\gamma > 10^{6}$, these terms convey an enormous $> 10^{48}$ enhancement. Even still, there needs the presence of fluctuating magnetic fields of neutron star strength, to overcome the small neutrino magnetic moment $\mu_{\nu}$, value given by the Particle Data Group \cite{PDG}
\begin{equation}
\mu_{\nu} < 0.32 \times 10^{-10} \mu_{B}
\end{equation}
where $\mu_{B}$ is the Bohr magneton. The Standard Model allows a massive Dirac neutrino to possess a tiny magnetic dipole moment \cite{Fu} $\simeq 3 \times 10^{-19} (m_{\nu} / eV) \mu_{B}$ where $m_{\nu}$ is the neutrino mass in eV. However, most extensions of the Standard Model predict anomalous magnetic dipole moments that approach the experimental limit given by Eq(44). For example, reference \cite{AA} gives an extension of the minimal supersymmetric standard model by including a vector-like leptonic generation which arises in many grand unfied theories and string models. This leads to neutrino magnetic dipole moments many orders of magnitude larger than the Standard Model prediction and a clear signal for new physics.

From Eq(42), as long as $\vec{\beta}$ is not parallel to the cosmological magnetic field $\vec{B}$, then $\frac{d \vec{\mu}}{dt}$ is dominated by the first term of Eq(42), for large fields
\begin{equation}
\dot{\vec{\mu}}^{2} \approx u^{2}\vec{\mu}^{2}\vec{B}^{2}
\end{equation}
Characterizing the fluctuating cosmological field by its impact on the neutrino by
\begin{equation}
\dot{\vec{\beta}} \cdot \vec{\beta} \approx \omega
\end{equation}
and making the parameterizations
\begin{equation}
|\vec{B}| = 10^{a} \;  {\rm  Tesla}, \; \mu_{\nu} = 10^{-x}\mu_{B}
\end{equation}
 with $\mu_{B}^{2} \; {\rm Tesla}^{2} = [5.788\times10^{-11} \; {\rm MeV}]^{2}$ then
\begin{eqnarray}
P_{\rm COSMOLOGICAL} & \approx &  \frac{2}{3 c^{3}} [\gamma^{8} (\vec{\beta}\cdot \dot{\vec{\beta}})^{2} \dot{\vec{\mu}}^{2}] \nonumber \\
    & \sim & 2.19 \;  \omega^{2} 10^{2a - 2x + 9}    \; {\rm MeV/s} 
\end{eqnarray}

There are certainly regions, Fig.\ 1,  of parameter space $\{a,x,\omega\}$ where the magnetic dipole moment radiation leads to fast neutrino cooling. 

\begin{figure}[htp]
\includegraphics[scale=0.25]{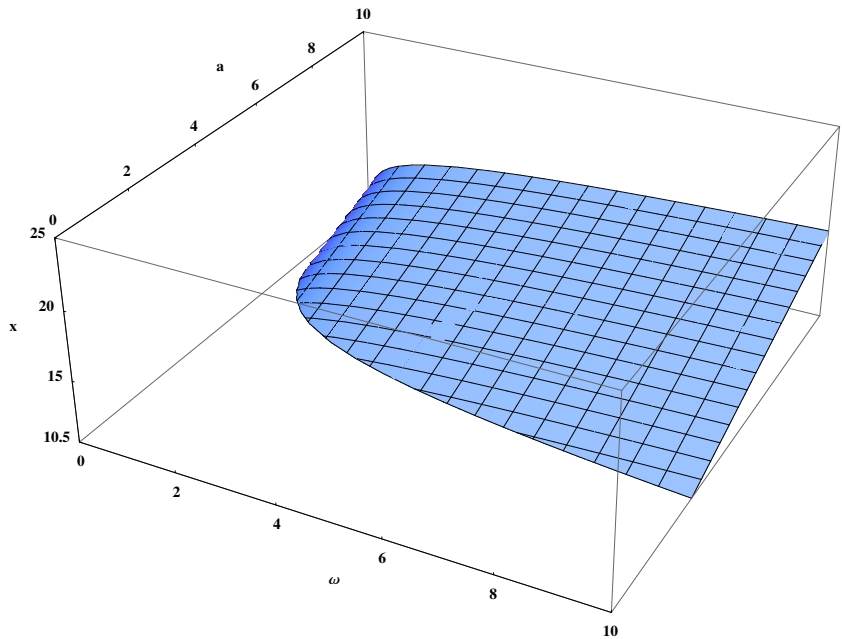}
\caption[*]{Parameter space for 1 MeV/s radiation emission.}
\label{Fig. 1}
\end{figure}

\section{Conclusion}
\hspace{0.25in}It can be argued successfully that any physics associated with magnetic moments is subtle: the electric charge is a Lorentz scalar, whereas the magnetic dipole moment is a significant component of a second order anti-symmetric tensor, ensuring complicated dynamics. It was well known that ordinary magnetic fields cannot lead to measurable magnetic moment radiation losses, but the enhancement for relativistic particles was not anticipated. 

Because cosmological neutrinos and anti-neutrinos are not expected to mutually annihilate in the expansion of the universe as do the charged particles, there would be an almost unlimited number of neutrinos and anti-neutrinos surviving the Big Bang. If these cosmological neutrinos and anti-neutrinos radiate energy after they decouple from matter, and if this loss is a significant fraction of their thermal energy, they can condense and be a component of dark matter. Condensed neutrino matter would then form the largest structures in the universe, dwarfing the visible galaxies in size and mass. This paper outlines a viable cooling mechanism that allows such a scenario.

Further work requires a detailed simulation of neutrino primordial density fluctuations in concert with a full electromagnetic treatment of early universe magnetic field generations in order to answer the question of neutrino condensation timelines.

\end{document}